%% file: samplepaper.tex
\documentclass[runningheads]{llncs}

\usepackage{graphicx}
\usepackage[T1]{fontenc}
\usepackage[utf8]{inputenc}
\usepackage{graphicx}
\usepackage{microtype}
\usepackage{cite}
\usepackage[hyphens]{url}
\usepackage{amssymb}
\usepackage{amsmath}
\usepackage[mathscr]{euscript}
\usepackage{multirow}
\usepackage{tabularx}
\usepackage{diagbox}
\usepackage{adjustbox}
\usepackage{subfigure}
\usepackage{tikz}
\usepackage{enumitem}

\usepackage[misc,geometry]{ifsym}
\usepackage{outlines}

\usetikzlibrary{fadings}
\usetikzlibrary{patterns}
\usetikzlibrary{shadows.blur}
\usetikzlibrary{shapes}

\usepackage[colorlinks = true, allcolors=blue]{hyperref}

\begin{document}

\title{CoVaxNet: An Online-Offline Data Repository for COVID-19 Vaccine Hesitancy Research}

\titlerunning{CoVaxNet: An Online-Offline COVID-19 Vaccine Data Repository}

\author{Bohan Jiang\textsuperscript{(\Letter)} \and
Paras Sheth \and 
Baoxin Li \and
Huan Liu}
\authorrunning{B. Jiang et al.}
% \author{Bohan Jiang\textsuperscript{(\Letter)}\orcidlink{0000-0001-8552-2681} \and
% Paras Sheth\orcidlink{0000-0002-6186-6946} \and 
% Baoxin Li \and
% Huan Liu\orcidlink{0000-0002-3264-7904}}
% %
% \authorrunning{B. Jiang et al.}
\institute{Arizona State University, Tempe AZ, USA \\
\email{\{bjiang14, psheth5, baoxin.li, huanliu\}@asu.edu}}
\maketitle              % typeset the header of the contribution
\begin{abstract}
\input{Abstract/abstract}
\end{abstract}

\vspace{-0.6cm}
\section{Introduction} \label{intro}
\input{Introduction/introduction}
\vspace{-0.3cm}
\section{Related Work} \label{relate}
\input{Related work/relate}

\section{Construction: What is in CoVaxNet?} \label{data}
\input{Data collection/data}

\section{Insight: What is in the Data?} \label{analysis}
\input{Analysis/analysis}

\section{Online and Offline Data Connection and Application} \label{connection}
\input{Connection/connection}

\section{Conclusion}
\input{Conclusion/conclusion}

% \section*{Acknowledgements}

\bibliographystyle{splncs04}
\bibliography{ref}
% \bibliography{bohan, lu, ref_echo, ref_echo_2}
%

\end{document}

%% file: Abstract/abstract.tex
%Despite the astonishing success of COVID-19 vaccines against the virus, a substantial proportion of the population is still hesitant to be vaccinated, undermining governmental efforts to control the virus. To mitigate this problem, we need to consider a wide-ranging set of factors such as social media discourses, news media propaganda, government responses, demographic and socioeconomic statuses, and COVID-19 statistics. In this paper, we present a multi-source, multi-modal, and multi-feature online-offline data repository \texttt{M\textsuperscript{3}O\textsuperscript{2}CoVax}, which is a compilation of various online and offline data that can be crucial for combating the COVID-19 vaccine hesitancy. We also provide a descriptive analysis of the repository to help researchers understand the features associated with each dataset and how to explore online-offline data encompassing the vaccine campaign.
Despite the astonishing success of COVID-19 vaccines against the virus, a substantial proportion of the population is still hesitant to be vaccinated, undermining governmental efforts to control the virus. To address this problem, we need to understand the different factors giving rise to such a behavior, including social media discourses, news media propaganda, government responses, demographic and socioeconomic statuses, and COVID-19 statistics, etc. However, existing datasets fail to cover all these aspects, making it difficult to form a complete picture in inferencing about the problem of vaccine hesitancy. In this paper, we construct a multi-source, multi-modal, and multi-feature online-offline data repository \texttt{CoVaxNet}\footnote{\url{https://github.com/jiangbohan/CoVaxNet}}. We provide descriptive analyses and insights to illustrate critical patterns in \texttt{CoVaxNet}. Moreover, we propose a novel approach for connecting online and offline data so as to facilitate the inference tasks that exploit complementary information sources.   % to facilitate researchers establishing relations between online and offline data. The connected online-offline data have the potential to benefit data-driven tasks.
%Such a relation can aid in answering important questions such as what are the crucial factors that influence people to be hesitant to the vaccines.

%% file: Introduction/introduction.tex
The COVID-19 pandemic has killed over six million people and infected 536 million globally as of mid-June, 2022\footnote{\url{https://covid19.who.int/}}. It has been pointed out that the FDA-authorized COVID-19 vaccines are highly effective at protecting against severe illness and reinfection\cite{kim2021looking}. However, according to a recent KFF survey\footnote{\url{https://www.kff.org/coronavirus-covid-19/dashboard/}}, 32\% of people in the U.S. %said they would \textit{``wait and see'', ``only if required'',} or \textit{``definitely not''}
showed hesitancy to receive at least one dose of the COVID-19 vaccine. SARS-CoV-2 is constantly mutating to highly contagious variants and subvariants, causing the number of cumulative cases of infection continues to grow in most countries. Therefore, the effort to end the pandemic will be severely hindered if a substantial proportion of the population still show \textit{vaccine hesitancy}.
%or \textit{vaccine refusal}.

Mitigating the ongoing COVID-19 vaccine hesitancy presents unique challenges. First, as defined in ~\cite{macdonald2015vaccine}, vaccine hesitancy is a continuum between full acceptance (pro-vaccine) and outright refusal of all vaccines (anti-vaccine), and  the COVID-19 vaccine hesitancy is driven by a complex set of dynamic context-specific factors, including social media influence, public stance and sentiment, structural inequality, risk of the disease, and trust toward government%\cite{piltch2022determinants}
, etc. All these make the characterization of COVID-19 vaccine hesitancy a non-trivial task. Second, %as discussed in the \textit{Related Work} section, academics have started collecting data to analyze COVID-19 vaccine hesitancy. However, 
the research community lacks comprehensive COVID-19 vaccine datasets containing adequate information to support effective analysis of the factors mentioned above and their impact on COVID-19 vaccine hesitancy. Given the complexity of the problem, a data repository capturing online-offline COVID-19 vaccine information would be critical to the analysis. %This serves as our motivation to build such a data repository.

In this paper, we present a multi-source, multi-modal, and multi-feature online-offline data repository \texttt{CoVaxNet}. As shwon in Figure~\ref{fig::outline}, it contains two online datasets: (i) a social media dataset and (ii) a fact-checking dataset; and four offline datasets: (i) COVID-19 statistics, (ii) U.S. Census Bureau data, (iii) government responses, and (iv) local news reports. To the best of our knowledge, we are the first to build such a diverse online-offline COVID-19 vaccine repository. We aim to continuously update this data repository with new sources and features, as well as maintain completeness. The main contributions of this work are:
\begin{itemize}
  \item[$\bullet$] We construct a multi-source, multi-modal, and multi-feature online-offline repository to facilitate COVID-19 vaccine hesitancy related research such as (i) detecting COVID-19 vaccine misinformation and stance, (ii) exploring the effect of online activities on offline outcomes and vice-versa, and (iii) finding correlations between structural inequality and vaccine hesitancy;
  \item[$\bullet$] We provide descriptive analyzes of the repository from different perspectives. We illustrate insights of various features and characteristics in each dataset, which covers textual, visual, spatio-temporal, and network information; and 
  \item[$\bullet$] We propose a new online and offline data connection approach with examples from \texttt{CoVaxNet} and demonstrate baseline performances for stance detection.
\end{itemize}
\input{figure/dataset_overview_big}

%% file: figure/dataset_overview_big.tex
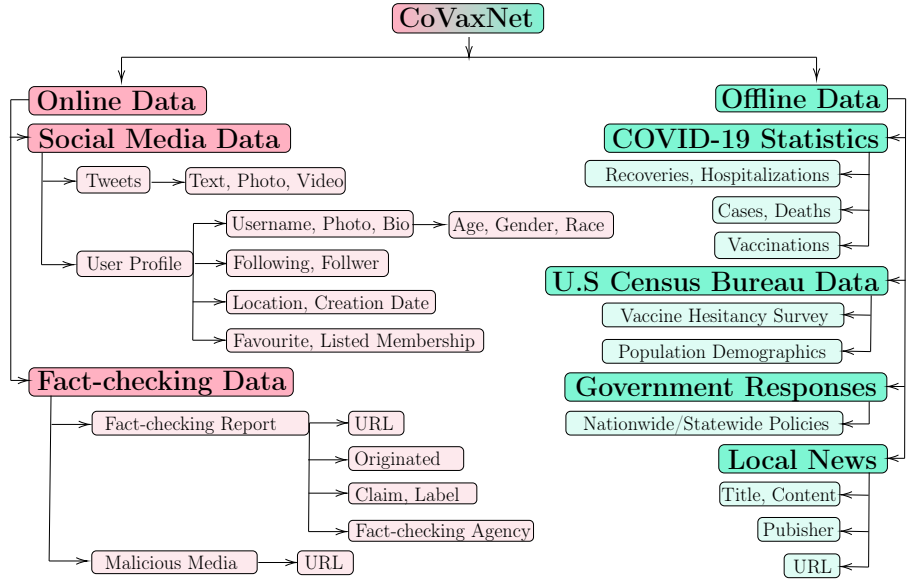
\begin{figure}
\centering
\resizebox{1\linewidth}{!}{

% Gradient Info
  
\tikzset {_3mrg93qid/.code = {\pgfsetadditionalshadetransform{ \pgftransformshift{\pgfpoint{0 bp } { 0 bp }  }  \pgftransformrotate{0 }  \pgftransformscale{2 }  }}}
\pgfdeclarehorizontalshading{_ai3icsslp}{150bp}{rgb(0bp)=(1,0.69,0.76);
rgb(37.5bp)=(1,0.69,0.76);
rgb(62.5bp)=(0.49,0.96,0.83);
rgb(100bp)=(0.49,0.96,0.83)}
\tikzset{every picture/.style={line width=0.75pt}} %set default line width to 0.75pt        

\begin{tikzpicture}[x=0.75pt,y=0.75pt,yscale=-1,xscale=1]
%uncomment if require: \path (0,861); %set diagram left start at 0, and has height of 861

%Rounded Rect [id:dp46519577549696134] 
\draw  [fill={rgb, 255:red, 255; green, 177; blue, 194 }  ,fill opacity=1 ] (243,132.6) .. controls (243,129.51) and (245.51,127) .. (248.6,127) -- (412.8,127) .. controls (415.89,127) and (418.4,129.51) .. (418.4,132.6) -- (418.4,149.4) .. controls (418.4,152.49) and (415.89,155) .. (412.8,155) -- (248.6,155) .. controls (245.51,155) and (243,152.49) .. (243,149.4) -- cycle ;
%Rounded Rect [id:dp5175443432657261] 
\path  [shading=_ai3icsslp,_3mrg93qid] (602,50.8) .. controls (602,47.6) and (604.6,45) .. (607.8,45) -- (745.6,45) .. controls (748.8,45) and (751.4,47.6) .. (751.4,50.8) -- (751.4,68.2) .. controls (751.4,71.4) and (748.8,74) .. (745.6,74) -- (607.8,74) .. controls (604.6,74) and (602,71.4) .. (602,68.2) -- cycle ; % for fading 
 \draw   (602,50.8) .. controls (602,47.6) and (604.6,45) .. (607.8,45) -- (745.6,45) .. controls (748.8,45) and (751.4,47.6) .. (751.4,50.8) -- (751.4,68.2) .. controls (751.4,71.4) and (748.8,74) .. (745.6,74) -- (607.8,74) .. controls (604.6,74) and (602,71.4) .. (602,68.2) -- cycle ; % for border 

%Rounded Rect [id:dp09917266670434377] 
\draw  [fill={rgb, 255:red, 126; green, 246; blue, 211 }  ,fill opacity=1 ] (921.4,130.8) .. controls (921.4,127.87) and (923.77,125.5) .. (926.7,125.5) -- (1085.7,125.5) .. controls (1088.63,125.5) and (1091,127.87) .. (1091,130.8) -- (1091,146.7) .. controls (1091,149.63) and (1088.63,152) .. (1085.7,152) -- (926.7,152) .. controls (923.77,152) and (921.4,149.63) .. (921.4,146.7) -- cycle ;
%Straight Lines [id:da6823632091600484] 
\draw    (334.4,98) -- (1021,98) ;
%Straight Lines [id:da016775810871046515] 
\draw    (334.4,98) -- (334.4,106.5) -- (334.4,120) ;
\draw [shift={(334.4,122)}, rotate = 270] [color={rgb, 255:red, 0; green, 0; blue, 0 }  ][line width=0.75]    (10.93,-3.29) .. controls (6.95,-1.4) and (3.31,-0.3) .. (0,0) .. controls (3.31,0.3) and (6.95,1.4) .. (10.93,3.29)   ;
%Straight Lines [id:da28124581887798583] 
\draw    (677.7,78) -- (677.7,96) ;
\draw [shift={(677.7,98)}, rotate = 270] [color={rgb, 255:red, 0; green, 0; blue, 0 }  ][line width=0.75]    (10.93,-3.29) .. controls (6.95,-1.4) and (3.31,-0.3) .. (0,0) .. controls (3.31,0.3) and (6.95,1.4) .. (10.93,3.29)   ;
%Straight Lines [id:da6284532413533035] 
\draw    (1021,98) -- (1021,118) ;
\draw [shift={(1021,120)}, rotate = 270] [color={rgb, 255:red, 0; green, 0; blue, 0 }  ][line width=0.75]    (10.93,-3.29) .. controls (6.95,-1.4) and (3.31,-0.3) .. (0,0) .. controls (3.31,0.3) and (6.95,1.4) .. (10.93,3.29)   ;
%Rounded Rect [id:dp3006631936400972] 
\draw  [fill={rgb, 255:red, 255; green, 177; blue, 194 }  ,fill opacity=1 ] (242,169.6) .. controls (242,166.51) and (244.51,164) .. (247.6,164) -- (495.8,164) .. controls (498.89,164) and (501.4,166.51) .. (501.4,169.6) -- (501.4,186.4) .. controls (501.4,189.49) and (498.89,192) .. (495.8,192) -- (247.6,192) .. controls (244.51,192) and (242,189.49) .. (242,186.4) -- cycle ;
%Rounded Rect [id:dp3622538673931921] 
\draw  [fill={rgb, 255:red, 255; green, 177; blue, 194 }  ,fill opacity=1 ] (243,410.6) .. controls (243,407.51) and (245.51,405) .. (248.6,405) -- (498.47,405) .. controls (501.56,405) and (504.07,407.51) .. (504.07,410.6) -- (504.07,427.4) .. controls (504.07,430.49) and (501.56,433) .. (498.47,433) -- (248.6,433) .. controls (245.51,433) and (243,430.49) .. (243,427.4) -- cycle ;
%Straight Lines [id:da36471000923713315] 
\draw    (225,140) -- (225,255) -- (225,417.5) ;
%Straight Lines [id:da5329232856395267] 
\draw    (225,140) -- (243,140) ;
%Straight Lines [id:da9635044502071781] 
\draw    (224,177) -- (240,177) ;
\draw [shift={(242,177)}, rotate = 180] [color={rgb, 255:red, 0; green, 0; blue, 0 }  ][line width=0.75]    (10.93,-3.29) .. controls (6.95,-1.4) and (3.31,-0.3) .. (0,0) .. controls (3.31,0.3) and (6.95,1.4) .. (10.93,3.29)   ;
%Straight Lines [id:da03273504536809124] 
\draw    (1108,494.5) -- (1109,139) ;
%Straight Lines [id:da38332624589518316] 
\draw    (256.62,220.5) -- (286.38,220.5) ;
\draw [shift={(288.38,220.5)}, rotate = 180] [color={rgb, 255:red, 0; green, 0; blue, 0 }  ][line width=0.75]    (10.93,-3.29) .. controls (6.95,-1.4) and (3.31,-0.3) .. (0,0) .. controls (3.31,0.3) and (6.95,1.4) .. (10.93,3.29)   ;
%Rounded Rect [id:dp1587868048346508] 
\draw  [fill={rgb, 255:red, 253; green, 233; blue, 237 }  ,fill opacity=1 ] (290.33,211.24) .. controls (290.33,208.35) and (292.68,206) .. (295.57,206) -- (357.15,206) .. controls (360.04,206) and (362.39,208.35) .. (362.39,211.24) -- (362.39,226.97) .. controls (362.39,229.87) and (360.04,232.21) .. (357.15,232.21) -- (295.57,232.21) .. controls (292.68,232.21) and (290.33,229.87) .. (290.33,226.97) -- cycle ;
%Rounded Rect [id:dp7974097163619687] 
\draw  [fill={rgb, 255:red, 253; green, 233; blue, 237 }  ,fill opacity=1 ] (290.33,294.68) .. controls (290.33,291.91) and (292.58,289.66) .. (295.35,289.66) -- (400.12,289.66) .. controls (402.89,289.66) and (405.14,291.91) .. (405.14,294.68) -- (405.14,309.74) .. controls (405.14,312.52) and (402.89,314.76) .. (400.12,314.76) -- (295.35,314.76) .. controls (292.58,314.76) and (290.33,312.52) .. (290.33,309.74) -- cycle ;
%Straight Lines [id:da25921912810206527] 
\draw    (256,192.5) -- (255.4,303.05) ;
%Straight Lines [id:da5662449740521915] 
\draw    (255.4,303.05) -- (285.16,303.05) ;
\draw [shift={(287.16,303.05)}, rotate = 180] [color={rgb, 255:red, 0; green, 0; blue, 0 }  ][line width=0.75]    (10.93,-3.29) .. controls (6.95,-1.4) and (3.31,-0.3) .. (0,0) .. controls (3.31,0.3) and (6.95,1.4) .. (10.93,3.29)   ;
%Straight Lines [id:da7642763932503138] 
\draw    (364.1,221.62) -- (393.86,221.62) ;
\draw [shift={(395.86,221.62)}, rotate = 180] [color={rgb, 255:red, 0; green, 0; blue, 0 }  ][line width=0.75]    (10.93,-3.29) .. controls (6.95,-1.4) and (3.31,-0.3) .. (0,0) .. controls (3.31,0.3) and (6.95,1.4) .. (10.93,3.29)   ;
%Rounded Rect [id:dp10868743457860619] 
\draw  [fill={rgb, 255:red, 253; green, 233; blue, 237 }  ,fill opacity=1 ] (397.81,212.36) .. controls (397.81,209.46) and (400.16,207.12) .. (403.05,207.12) -- (550.12,207.12) .. controls (553.02,207.12) and (555.37,209.46) .. (555.37,212.36) -- (555.37,228.09) .. controls (555.37,230.98) and (553.02,233.33) .. (550.12,233.33) -- (403.05,233.33) .. controls (400.16,233.33) and (397.81,230.98) .. (397.81,228.09) -- cycle ;
%Straight Lines [id:da8790111809282946] 
\draw    (406.85,301.94) -- (436.6,301.94) ;
\draw [shift={(438.6,301.94)}, rotate = 180] [color={rgb, 255:red, 0; green, 0; blue, 0 }  ][line width=0.75]    (10.93,-3.29) .. controls (6.95,-1.4) and (3.31,-0.3) .. (0,0) .. controls (3.31,0.3) and (6.95,1.4) .. (10.93,3.29)   ;
%Rounded Rect [id:dp8835004684396743] 
\draw  [fill={rgb, 255:red, 253; green, 233; blue, 237 }  ,fill opacity=1 ] (438.12,293.79) .. controls (438.12,290.9) and (440.46,288.55) .. (443.36,288.55) -- (592.87,288.55) .. controls (595.77,288.55) and (598.11,290.9) .. (598.11,293.79) -- (598.11,309.52) .. controls (598.11,312.42) and (595.77,314.76) .. (592.87,314.76) -- (443.36,314.76) .. controls (440.46,314.76) and (438.12,312.42) .. (438.12,309.52) -- cycle ;
%Straight Lines [id:da8955187678425327] 
\draw    (405.5,262.33) -- (404.77,377.74) ;
%Straight Lines [id:da7781765164603622] 
\draw    (405.5,262.33) -- (435.26,262.33) ;
\draw [shift={(437.26,262.33)}, rotate = 180] [color={rgb, 255:red, 0; green, 0; blue, 0 }  ][line width=0.75]    (10.93,-3.29) .. controls (6.95,-1.4) and (3.31,-0.3) .. (0,0) .. controls (3.31,0.3) and (6.95,1.4) .. (10.93,3.29)   ;
%Straight Lines [id:da5294498978113469] 
\draw    (404.77,339.81) -- (434.53,339.81) ;
\draw [shift={(436.53,339.81)}, rotate = 180] [color={rgb, 255:red, 0; green, 0; blue, 0 }  ][line width=0.75]    (10.93,-3.29) .. controls (6.95,-1.4) and (3.31,-0.3) .. (0,0) .. controls (3.31,0.3) and (6.95,1.4) .. (10.93,3.29)   ;
%Rounded Rect [id:dp05880133048263869] 
\draw  [fill={rgb, 255:red, 253; green, 233; blue, 237 }  ,fill opacity=1 ] (438.12,254.75) .. controls (438.12,251.85) and (440.46,249.51) .. (443.36,249.51) -- (617.3,249.51) .. controls (620.19,249.51) and (622.54,251.85) .. (622.54,254.75) -- (622.54,270.48) .. controls (622.54,273.37) and (620.19,275.72) .. (617.3,275.72) -- (443.36,275.72) .. controls (440.46,275.72) and (438.12,273.37) .. (438.12,270.48) -- cycle ;
%Rounded Rect [id:dp43341182472757356] 
\draw  [fill={rgb, 255:red, 253; green, 233; blue, 237 }  ,fill opacity=1 ] (438.12,333.14) .. controls (438.12,330.59) and (440.18,328.53) .. (442.72,328.53) -- (639.92,328.53) .. controls (642.46,328.53) and (644.52,330.59) .. (644.52,333.14) -- (644.52,346.97) .. controls (644.52,349.51) and (642.46,351.58) .. (639.92,351.58) -- (442.72,351.58) .. controls (440.18,351.58) and (438.12,349.51) .. (438.12,346.97) -- cycle ;
%Straight Lines [id:da13336864293850104] 
\draw    (404.77,377.74) -- (434.53,377.74) ;
\draw [shift={(436.53,377.74)}, rotate = 180] [color={rgb, 255:red, 0; green, 0; blue, 0 }  ][line width=0.75]    (10.93,-3.29) .. controls (6.95,-1.4) and (3.31,-0.3) .. (0,0) .. controls (3.31,0.3) and (6.95,1.4) .. (10.93,3.29)   ;
%Rounded Rect [id:dp22666948607233728] 
\draw  [fill={rgb, 255:red, 253; green, 233; blue, 237 }  ,fill opacity=1 ] (438.12,371.07) .. controls (438.12,368.52) and (440.18,366.46) .. (442.72,366.46) -- (687.55,366.46) .. controls (690.09,366.46) and (692.16,368.52) .. (692.16,371.07) -- (692.16,384.9) .. controls (692.16,387.44) and (690.09,389.5) .. (687.55,389.5) -- (442.72,389.5) .. controls (440.18,389.5) and (438.12,387.44) .. (438.12,384.9) -- cycle ;
%Straight Lines [id:da8269624465914651] 
\draw    (225,417.5) -- (241,417.5) ;
\draw [shift={(243,417.5)}, rotate = 180] [color={rgb, 255:red, 0; green, 0; blue, 0 }  ][line width=0.75]    (10.93,-3.29) .. controls (6.95,-1.4) and (3.31,-0.3) .. (0,0) .. controls (3.31,0.3) and (6.95,1.4) .. (10.93,3.29)   ;
%Straight Lines [id:da30874748842082655] 
\draw    (621.69,263.45) -- (651.44,263.45) ;
\draw [shift={(653.44,263.45)}, rotate = 180] [color={rgb, 255:red, 0; green, 0; blue, 0 }  ][line width=0.75]    (10.93,-3.29) .. controls (6.95,-1.4) and (3.31,-0.3) .. (0,0) .. controls (3.31,0.3) and (6.95,1.4) .. (10.93,3.29)   ;
%Rounded Rect [id:dp8344524247354315] 
\draw  [fill={rgb, 255:red, 253; green, 233; blue, 237 }  ,fill opacity=1 ] (657.96,255.86) .. controls (657.96,252.97) and (660.31,250.62) .. (663.2,250.62) -- (815.16,250.62) .. controls (818.05,250.62) and (820.4,252.97) .. (820.4,255.86) -- (820.4,271.59) .. controls (820.4,274.49) and (818.05,276.84) .. (815.16,276.84) -- (663.2,276.84) .. controls (660.31,276.84) and (657.96,274.49) .. (657.96,271.59) -- cycle ;
%Straight Lines [id:da41090559384231007] 
\draw    (265.34,461) -- (301.05,461) ;
\draw [shift={(303.05,461)}, rotate = 180] [color={rgb, 255:red, 0; green, 0; blue, 0 }  ][line width=0.75]    (10.93,-3.29) .. controls (6.95,-1.4) and (3.31,-0.3) .. (0,0) .. controls (3.31,0.3) and (6.95,1.4) .. (10.93,3.29)   ;
%Rounded Rect [id:dp16047019778393734] 
\draw  [fill={rgb, 255:red, 253; green, 233; blue, 237 }  ,fill opacity=1 ] (305.37,452.7) .. controls (305.37,450.1) and (307.47,448) .. (310.07,448) -- (514.7,448) .. controls (517.3,448) and (519.4,450.1) .. (519.4,452.7) -- (519.4,466.8) .. controls (519.4,469.4) and (517.3,471.5) .. (514.7,471.5) -- (310.07,471.5) .. controls (307.47,471.5) and (305.37,469.4) .. (305.37,466.8) -- cycle ;
%Rounded Rect [id:dp15281848946088483] 
\draw  [fill={rgb, 255:red, 253; green, 233; blue, 237 }  ,fill opacity=1 ] (305.02,590.5) .. controls (305.02,588.01) and (307.03,586) .. (309.52,586) -- (463.9,586) .. controls (466.39,586) and (468.4,588.01) .. (468.4,590.5) -- (468.4,604) .. controls (468.4,606.49) and (466.39,608.5) .. (463.9,608.5) -- (309.52,608.5) .. controls (307.03,608.5) and (305.02,606.49) .. (305.02,604) -- cycle ;
%Straight Lines [id:da5737361446587657] 
\draw    (264.76,432.5) -- (262.89,596) ;
%Straight Lines [id:da6558681810745599] 
\draw    (520.06,459.5) -- (519.19,566.95) ;
%Straight Lines [id:da7836338270472416] 
\draw    (518.76,496) -- (554.47,496) ;
\draw [shift={(556.47,496)}, rotate = 180] [color={rgb, 255:red, 0; green, 0; blue, 0 }  ][line width=0.75]    (10.93,-3.29) .. controls (6.95,-1.4) and (3.31,-0.3) .. (0,0) .. controls (3.31,0.3) and (6.95,1.4) .. (10.93,3.29)   ;
%Rounded Rect [id:dp4450761222005213] 
\draw  [fill={rgb, 255:red, 253; green, 233; blue, 237 }  ,fill opacity=1 ] (558.79,487.7) .. controls (558.79,485.1) and (560.89,483) .. (563.49,483) -- (667.7,483) .. controls (670.3,483) and (672.4,485.1) .. (672.4,487.7) -- (672.4,501.8) .. controls (672.4,504.4) and (670.3,506.5) .. (667.7,506.5) -- (563.49,506.5) .. controls (560.89,506.5) and (558.79,504.4) .. (558.79,501.8) -- cycle ;
%Straight Lines [id:da7264459135853614] 
\draw    (520.06,459.5) -- (555.77,459.5) ;
\draw [shift={(557.77,459.5)}, rotate = 180] [color={rgb, 255:red, 0; green, 0; blue, 0 }  ][line width=0.75]    (10.93,-3.29) .. controls (6.95,-1.4) and (3.31,-0.3) .. (0,0) .. controls (3.31,0.3) and (6.95,1.4) .. (10.93,3.29)   ;
%Straight Lines [id:da2534442585474821] 
\draw    (519.19,528.95) -- (554.9,528.95) ;
\draw [shift={(556.9,528.95)}, rotate = 180] [color={rgb, 255:red, 0; green, 0; blue, 0 }  ][line width=0.75]    (10.93,-3.29) .. controls (6.95,-1.4) and (3.31,-0.3) .. (0,0) .. controls (3.31,0.3) and (6.95,1.4) .. (10.93,3.29)   ;
%Rounded Rect [id:dp10396896466440553] 
\draw  [fill={rgb, 255:red, 253; green, 233; blue, 237 }  ,fill opacity=1 ] (558.79,452.7) .. controls (558.79,450.1) and (560.89,448) .. (563.49,448) -- (608.7,448) .. controls (611.3,448) and (613.4,450.1) .. (613.4,452.7) -- (613.4,466.8) .. controls (613.4,469.4) and (611.3,471.5) .. (608.7,471.5) -- (563.49,471.5) .. controls (560.89,471.5) and (558.79,469.4) .. (558.79,466.8) -- cycle ;
%Rounded Rect [id:dp6050913038696024] 
\draw  [fill={rgb, 255:red, 253; green, 233; blue, 237 }  ,fill opacity=1 ] (558.79,522.97) .. controls (558.79,520.69) and (560.64,518.84) .. (562.92,518.84) -- (681.27,518.84) .. controls (683.55,518.84) and (685.4,520.69) .. (685.4,522.97) -- (685.4,535.37) .. controls (685.4,537.65) and (683.55,539.5) .. (681.27,539.5) -- (562.92,539.5) .. controls (560.64,539.5) and (558.79,537.65) .. (558.79,535.37) -- cycle ;
%Straight Lines [id:da3512455207127225] 
\draw    (519.19,566.95) -- (554.9,566.95) ;
\draw [shift={(556.9,566.95)}, rotate = 180] [color={rgb, 255:red, 0; green, 0; blue, 0 }  ][line width=0.75]    (10.93,-3.29) .. controls (6.95,-1.4) and (3.31,-0.3) .. (0,0) .. controls (3.31,0.3) and (6.95,1.4) .. (10.93,3.29)   ;
%Straight Lines [id:da6000944372808876] 
\draw    (262.89,596) -- (298.6,596) ;
\draw [shift={(300.6,596)}, rotate = 180] [color={rgb, 255:red, 0; green, 0; blue, 0 }  ][line width=0.75]    (10.93,-3.29) .. controls (6.95,-1.4) and (3.31,-0.3) .. (0,0) .. controls (3.31,0.3) and (6.95,1.4) .. (10.93,3.29)   ;
%Straight Lines [id:da9731714992473457] 
\draw    (1091,139) -- (1109,139) ;
%Rounded Rect [id:dp29072084036775236] 
\draw  [fill={rgb, 255:red, 126; green, 246; blue, 211 }  ,fill opacity=1 ] (811.4,169.6) .. controls (811.4,166.51) and (813.91,164) .. (817,164) -- (1085.4,164) .. controls (1088.49,164) and (1091,166.51) .. (1091,169.6) -- (1091,186.4) .. controls (1091,189.49) and (1088.49,192) .. (1085.4,192) -- (817,192) .. controls (813.91,192) and (811.4,189.49) .. (811.4,186.4) -- cycle ;
%Straight Lines [id:da08782227580143487] 
\draw    (1109,177.5) -- (1094,177.5) ;
\draw [shift={(1092,177.5)}, rotate = 360] [color={rgb, 255:red, 0; green, 0; blue, 0 }  ][line width=0.75]    (10.93,-3.29) .. controls (6.95,-1.4) and (3.31,-0.3) .. (0,0) .. controls (3.31,0.3) and (6.95,1.4) .. (10.93,3.29)   ;
%Straight Lines [id:da03933839035211206] 
\draw    (1073,191.5) -- (1072.4,284.66) ;
%Straight Lines [id:da036686787872568294] 
\draw    (1073.4,214) -- (1047,213.63) ;
\draw [shift={(1045,213.6)}, rotate = 0.8] [color={rgb, 255:red, 0; green, 0; blue, 0 }  ][line width=0.75]    (10.93,-3.29) .. controls (6.95,-1.4) and (3.31,-0.3) .. (0,0) .. controls (3.31,0.3) and (6.95,1.4) .. (10.93,3.29)   ;
%Rounded Rect [id:dp6921404238205919] 
\draw  [fill={rgb, 255:red, 126; green, 246; blue, 211 }  ,fill opacity=0.25 ] (792.4,205.22) .. controls (792.4,202.34) and (794.74,200) .. (797.62,200) -- (1038.78,200) .. controls (1041.66,200) and (1044,202.34) .. (1044,205.22) -- (1044,220.89) .. controls (1044,223.77) and (1041.66,226.11) .. (1038.78,226.11) -- (797.62,226.11) .. controls (794.74,226.11) and (792.4,223.77) .. (792.4,220.89) -- cycle ;
%Straight Lines [id:da0040899186133616094] 
\draw    (1073.4,249.32) -- (1048,248.81) ;
\draw [shift={(1046,248.77)}, rotate = 1.14] [color={rgb, 255:red, 0; green, 0; blue, 0 }  ][line width=0.75]    (10.93,-3.29) .. controls (6.95,-1.4) and (3.31,-0.3) .. (0,0) .. controls (3.31,0.3) and (6.95,1.4) .. (10.93,3.29)   ;
%Rounded Rect [id:dp09682937383606816] 
\draw  [fill={rgb, 255:red, 126; green, 246; blue, 211 }  ,fill opacity=0.25 ] (918.4,241.13) .. controls (918.4,238.25) and (920.74,235.91) .. (923.62,235.91) -- (1039.78,235.91) .. controls (1042.66,235.91) and (1045,238.25) .. (1045,241.13) -- (1045,256.8) .. controls (1045,259.68) and (1042.66,262.02) .. (1039.78,262.02) -- (923.62,262.02) .. controls (920.74,262.02) and (918.4,259.68) .. (918.4,256.8) -- cycle ;
%Straight Lines [id:da8103184507723242] 
\draw    (1072.4,284.66) -- (1047,284.16) ;
\draw [shift={(1045,284.12)}, rotate = 1.14] [color={rgb, 255:red, 0; green, 0; blue, 0 }  ][line width=0.75]    (10.93,-3.29) .. controls (6.95,-1.4) and (3.31,-0.3) .. (0,0) .. controls (3.31,0.3) and (6.95,1.4) .. (10.93,3.29)   ;
%Rounded Rect [id:dp36194664755051065] 
\draw  [fill={rgb, 255:red, 126; green, 246; blue, 211 }  ,fill opacity=0.25 ] (920.4,276.11) .. controls (920.4,273.23) and (922.74,270.89) .. (925.62,270.89) -- (1038.78,270.89) .. controls (1041.66,270.89) and (1044,273.23) .. (1044,276.11) -- (1044,291.78) .. controls (1044,294.66) and (1041.66,297) .. (1038.78,297) -- (925.62,297) .. controls (922.74,297) and (920.4,294.66) .. (920.4,291.78) -- cycle ;
%Rounded Rect [id:dp9069733526070529] 
\draw  [fill={rgb, 255:red, 126; green, 246; blue, 211 }  ,fill opacity=1 ] (753.4,310.6) .. controls (753.4,307.51) and (755.91,305) .. (759,305) -- (1085.84,305) .. controls (1088.94,305) and (1091.44,307.51) .. (1091.44,310.6) -- (1091.44,327.4) .. controls (1091.44,330.49) and (1088.94,333) .. (1085.84,333) -- (759,333) .. controls (755.91,333) and (753.4,330.49) .. (753.4,327.4) -- cycle ;
%Straight Lines [id:da3782167144367412] 
\draw    (1109,318.5) -- (1094,318.5) ;
\draw [shift={(1092,318.5)}, rotate = 360] [color={rgb, 255:red, 0; green, 0; blue, 0 }  ][line width=0.75]    (10.93,-3.29) .. controls (6.95,-1.4) and (3.31,-0.3) .. (0,0) .. controls (3.31,0.3) and (6.95,1.4) .. (10.93,3.29)   ;
%Straight Lines [id:da4494320135274119] 
\draw    (1075,333.5) -- (1074.4,389) ;
%Straight Lines [id:da957206952772568] 
\draw    (1074.4,353) -- (1049,352.54) ;
\draw [shift={(1047,352.5)}, rotate = 1.05] [color={rgb, 255:red, 0; green, 0; blue, 0 }  ][line width=0.75]    (10.93,-3.29) .. controls (6.95,-1.4) and (3.31,-0.3) .. (0,0) .. controls (3.31,0.3) and (6.95,1.4) .. (10.93,3.29)   ;
%Rounded Rect [id:dp49144947332678823] 
\draw  [fill={rgb, 255:red, 126; green, 246; blue, 211 }  ,fill opacity=0.25 ] (809.4,381.7) .. controls (809.4,379.1) and (811.5,377) .. (814.1,377) -- (1041.3,377) .. controls (1043.9,377) and (1046,379.1) .. (1046,381.7) -- (1046,395.8) .. controls (1046,398.4) and (1043.9,400.5) .. (1041.3,400.5) -- (814.1,400.5) .. controls (811.5,400.5) and (809.4,398.4) .. (809.4,395.8) -- cycle ;
%Straight Lines [id:da030268139896643786] 
\draw    (1074.4,389) -- (1049,388.54) ;
\draw [shift={(1047,388.5)}, rotate = 1.05] [color={rgb, 255:red, 0; green, 0; blue, 0 }  ][line width=0.75]    (10.93,-3.29) .. controls (6.95,-1.4) and (3.31,-0.3) .. (0,0) .. controls (3.31,0.3) and (6.95,1.4) .. (10.93,3.29)   ;
%Rounded Rect [id:dp14073006737244076] 
\draw  [fill={rgb, 255:red, 126; green, 246; blue, 211 }  ,fill opacity=0.25 ] (809.4,345.7) .. controls (809.4,343.1) and (811.5,341) .. (814.1,341) -- (1043.3,341) .. controls (1045.9,341) and (1048,343.1) .. (1048,345.7) -- (1048,359.8) .. controls (1048,362.4) and (1045.9,364.5) .. (1043.3,364.5) -- (814.1,364.5) .. controls (811.5,364.5) and (809.4,362.4) .. (809.4,359.8) -- cycle ;
%Rounded Rect [id:dp8725899443295015] 
\draw  [fill={rgb, 255:red, 126; green, 246; blue, 211 }  ,fill opacity=1 ] (772.4,415.6) .. controls (772.4,412.51) and (774.91,410) .. (778,410) -- (1084.84,410) .. controls (1087.94,410) and (1090.44,412.51) .. (1090.44,415.6) -- (1090.44,432.4) .. controls (1090.44,435.49) and (1087.94,438) .. (1084.84,438) -- (778,438) .. controls (774.91,438) and (772.4,435.49) .. (772.4,432.4) -- cycle ;
%Straight Lines [id:da17567275073896416] 
\draw    (1108,423.5) -- (1093,423.5) ;
\draw [shift={(1091,423.5)}, rotate = 360] [color={rgb, 255:red, 0; green, 0; blue, 0 }  ][line width=0.75]    (10.93,-3.29) .. controls (6.95,-1.4) and (3.31,-0.3) .. (0,0) .. controls (3.31,0.3) and (6.95,1.4) .. (10.93,3.29)   ;
%Straight Lines [id:da432041598951064] 
\draw    (1073,438.5) -- (1073,460.5) ;
%Rounded Rect [id:dp9649889121847566] 
\draw  [fill={rgb, 255:red, 126; green, 246; blue, 211 }  ,fill opacity=0.25 ] (773.4,452.7) .. controls (773.4,450.1) and (775.5,448) .. (778.1,448) -- (1042.3,448) .. controls (1044.9,448) and (1047,450.1) .. (1047,452.7) -- (1047,466.8) .. controls (1047,469.4) and (1044.9,471.5) .. (1042.3,471.5) -- (778.1,471.5) .. controls (775.5,471.5) and (773.4,469.4) .. (773.4,466.8) -- cycle ;
%Straight Lines [id:da3461498921075128] 
\draw    (469.5,597.5) -- (505.21,597.5) ;
\draw [shift={(507.21,597.5)}, rotate = 180] [color={rgb, 255:red, 0; green, 0; blue, 0 }  ][line width=0.75]    (10.93,-3.29) .. controls (6.95,-1.4) and (3.31,-0.3) .. (0,0) .. controls (3.31,0.3) and (6.95,1.4) .. (10.93,3.29)   ;
%Rounded Rect [id:dp12767786496974165] 
\draw  [fill={rgb, 255:red, 253; green, 233; blue, 237 }  ,fill opacity=1 ] (508.22,590.7) .. controls (508.22,588.1) and (510.33,586) .. (512.92,586) -- (558.7,586) .. controls (561.3,586) and (563.4,588.1) .. (563.4,590.7) -- (563.4,604.8) .. controls (563.4,607.4) and (561.3,609.5) .. (558.7,609.5) -- (512.92,609.5) .. controls (510.33,609.5) and (508.22,607.4) .. (508.22,604.8) -- cycle ;
%Rounded Rect [id:dp8789088247325461] 
\draw  [fill={rgb, 255:red, 126; green, 246; blue, 211 }  ,fill opacity=1 ] (924.4,486.6) .. controls (924.4,483.51) and (926.91,481) .. (930,481) -- (1085.4,481) .. controls (1088.49,481) and (1091,483.51) .. (1091,486.6) -- (1091,503.4) .. controls (1091,506.49) and (1088.49,509) .. (1085.4,509) -- (930,509) .. controls (926.91,509) and (924.4,506.49) .. (924.4,503.4) -- cycle ;
%Straight Lines [id:da6276017879239788] 
\draw    (1108,494.5) -- (1093,494.5) ;
\draw [shift={(1091,494.5)}, rotate = 360] [color={rgb, 255:red, 0; green, 0; blue, 0 }  ][line width=0.75]    (10.93,-3.29) .. controls (6.95,-1.4) and (3.31,-0.3) .. (0,0) .. controls (3.31,0.3) and (6.95,1.4) .. (10.93,3.29)   ;
%Straight Lines [id:da579506139241933] 
\draw    (1073,508.5) -- (1072.4,602) ;
%Straight Lines [id:da35033767548090045] 
\draw    (1072.4,531) -- (1047,530.54) ;
\draw [shift={(1045,530.5)}, rotate = 1.05] [color={rgb, 255:red, 0; green, 0; blue, 0 }  ][line width=0.75]    (10.93,-3.29) .. controls (6.95,-1.4) and (3.31,-0.3) .. (0,0) .. controls (3.31,0.3) and (6.95,1.4) .. (10.93,3.29)   ;
%Rounded Rect [id:dp7252638881904789] 
\draw  [fill={rgb, 255:red, 126; green, 246; blue, 211 }  ,fill opacity=0.25 ] (925.4,522.7) .. controls (925.4,520.1) and (927.5,518) .. (930.1,518) -- (1039.3,518) .. controls (1041.9,518) and (1044,520.1) .. (1044,522.7) -- (1044,536.8) .. controls (1044,539.4) and (1041.9,541.5) .. (1039.3,541.5) -- (930.1,541.5) .. controls (927.5,541.5) and (925.4,539.4) .. (925.4,536.8) -- cycle ;
%Straight Lines [id:da20297143791823458] 
\draw    (1072.4,567) -- (1047,566.54) ;
\draw [shift={(1045,566.5)}, rotate = 1.05] [color={rgb, 255:red, 0; green, 0; blue, 0 }  ][line width=0.75]    (10.93,-3.29) .. controls (6.95,-1.4) and (3.31,-0.3) .. (0,0) .. controls (3.31,0.3) and (6.95,1.4) .. (10.93,3.29)   ;
%Rounded Rect [id:dp6275645694538257] 
\draw  [fill={rgb, 255:red, 126; green, 246; blue, 211 }  ,fill opacity=0.25 ] (963,557.7) .. controls (963,555.1) and (965.1,553) .. (967.7,553) -- (1039.3,553) .. controls (1041.9,553) and (1044,555.1) .. (1044,557.7) -- (1044,571.8) .. controls (1044,574.4) and (1041.9,576.5) .. (1039.3,576.5) -- (967.7,576.5) .. controls (965.1,576.5) and (963,574.4) .. (963,571.8) -- cycle ;
%Straight Lines [id:da8277505937635605] 
\draw    (1072.4,602) -- (1047,601.54) ;
\draw [shift={(1045,601.5)}, rotate = 1.05] [color={rgb, 255:red, 0; green, 0; blue, 0 }  ][line width=0.75]    (10.93,-3.29) .. controls (6.95,-1.4) and (3.31,-0.3) .. (0,0) .. controls (3.31,0.3) and (6.95,1.4) .. (10.93,3.29)   ;
%Rounded Rect [id:dp15243891826209155] 
\draw  [fill={rgb, 255:red, 126; green, 246; blue, 211 }  ,fill opacity=0.25 ] (989,593.7) .. controls (989,591.1) and (991.1,589) .. (993.7,589) -- (1039.3,589) .. controls (1041.9,589) and (1044,591.1) .. (1044,593.7) -- (1044,607.8) .. controls (1044,610.4) and (1041.9,612.5) .. (1039.3,612.5) -- (993.7,612.5) .. controls (991.1,612.5) and (989,610.4) .. (989,607.8) -- cycle ;
%Straight Lines [id:da5488472545411058] 
\draw    (1073,460.5) -- (1048,460.5) ;
\draw [shift={(1046,460.5)}, rotate = 360] [color={rgb, 255:red, 0; green, 0; blue, 0 }  ][line width=0.75]    (10.93,-3.29) .. controls (6.95,-1.4) and (3.31,-0.3) .. (0,0) .. controls (3.31,0.3) and (6.95,1.4) .. (10.93,3.29)   ;
%Rounded Rect [id:dp7592779470907673] 
\draw  [fill={rgb, 255:red, 253; green, 233; blue, 237 }  ,fill opacity=1 ] (558.79,559.97) .. controls (558.79,557.69) and (560.64,555.84) .. (562.92,555.84) -- (737.27,555.84) .. controls (739.55,555.84) and (741.4,557.69) .. (741.4,559.97) -- (741.4,572.37) .. controls (741.4,574.65) and (739.55,576.5) .. (737.27,576.5) -- (562.92,576.5) .. controls (560.64,576.5) and (558.79,574.65) .. (558.79,572.37) -- cycle ;

% Text Node
\draw (251,167.6) node [anchor=north west][inner sep=0.75pt]   [align=left] 
{{\huge \textbf{Social Media Data}}};
% Text Node
\draw (242,406.6) node [anchor=north west][inner sep=0.75pt]   [align=left] {\begin{minipage}[lt]{196.25pt}\setlength\topsep{0pt}
\begin{center}
{\huge \textbf{Fact-checking Data}}
\end{center}

\end{minipage}};
% Text Node
\draw (249,130) node [anchor=north west][inner sep=0.75pt]   [align=left] {{\huge \textbf{Online Data}}};
% Text Node
\draw (292.33,211.24) node [anchor=north west][inner sep=0.75pt]   [align=left] {\begin{minipage}[lt]{49.26pt}\setlength\topsep{0pt}
\begin{center}
{\Large Tweets}
\end{center}

\end{minipage}};
% Text Node
\draw (292.33,293.68) node [anchor=north west][inner sep=0.75pt]   [align=left] {\begin{minipage}[lt]{79.45pt}\setlength\topsep{0pt}
\begin{center}
{\Large User Profile}
\end{center}

\end{minipage}};
% Text Node
\draw (394.81,211.36) node [anchor=north west][inner sep=0.75pt]   [align=left] {\begin{minipage}[lt]{121.67pt}\setlength\topsep{0pt}
\begin{center}
{\Large Text, Photo, Video}
\end{center}

\end{minipage}};
% Text Node
\draw (435.12,292.79) node [anchor=north west][inner sep=0.75pt]   [align=left] {\begin{minipage}[lt]{120.27pt}\setlength\topsep{0pt}
\begin{center}
{\Large Following, Follwer}
\end{center}

\end{minipage}};
% Text Node
\draw (432.12,253.75) node [anchor=north west][inner sep=0.75pt]   [align=left] {\begin{minipage}[lt]{146.41pt}\setlength\topsep{0pt}
\begin{center}
{\Large Username, Photo, Bio}
\end{center}

\end{minipage}};
% Text Node
\draw (435.12,332.14) node [anchor=north west][inner sep=0.75pt]   [align=left] {\begin{minipage}[lt]{157.04pt}\setlength\topsep{0pt}
\begin{center}
{\Large Location, Creation Date}
\end{center}

\end{minipage}};
% Text Node
\draw (432.12,370.07) node [anchor=north west][inner sep=0.75pt]   [align=left] {\begin{minipage}[lt]{196.2pt}\setlength\topsep{0pt}
\begin{center}
{\Large Favourite, Listed Membership}
\end{center}

\end{minipage}};
% Text Node
\draw (649.96,254.86) node [anchor=north west][inner sep=0.75pt]   [align=left] {\begin{minipage}[lt]{127.65pt}\setlength\topsep{0pt}
\begin{center}
{\Large Age, Gender, Race}
\end{center}

\end{minipage}};
% Text Node
\draw (306.37,451.7) node [anchor=north west][inner sep=0.75pt]   [align=left] {\begin{minipage}[lt]{142.32pt}\setlength\topsep{0pt}
\begin{center}
{\Large Fact-checking Report}
\end{center}

\end{minipage}};
% Text Node
\draw (310.02,589.5) node [anchor=north west][inner sep=0.75pt]   [align=left] {\begin{minipage}[lt]{108.03pt}\setlength\topsep{0pt}
\begin{center}
{\Large Malicious Media}
\end{center}

\end{minipage}};
% Text Node
\draw (558.79,486.7) node [anchor=north west][inner sep=0.75pt]   [align=left] {\begin{minipage}[lt]{70.49pt}\setlength\topsep{0pt}
\begin{center}
{\Large Originated}
\end{center}

\end{minipage}};
% Text Node
\draw (560.79,451.7) node [anchor=north west][inner sep=0.75pt]   [align=left] {\begin{minipage}[lt]{32.1pt}\setlength\topsep{0pt}
\begin{center}
{\Large URL}
\end{center}

\end{minipage}};
% Text Node
\draw (561.79,521.5) node [anchor=north west][inner sep=0.75pt]   [align=left] {\begin{minipage}[lt]{80.28pt}\setlength\topsep{0pt}
\begin{center}
{\Large Claim, Label}
\end{center}

\end{minipage}};
% Text Node
\draw (552.79,557.97) node [anchor=north west][inner sep=0.75pt]   [align=left] {\begin{minipage}[lt]{146.41pt}\setlength\topsep{0pt}
\begin{center}
{\Large Fact-checking Agency}
\end{center}

\end{minipage}};
% Text Node
\draw (925.4,127.8) node [anchor=north west][inner sep=0.75pt]   [align=left] {{\huge \textbf{Offline Data}}};
% Text Node
\draw (817.4,167.6) node [anchor=north west][inner sep=0.75pt]   [align=left] {\textbf{{\huge COVID-19 Statistics}}};
% Text Node
\draw (914.4,241.13) node [anchor=north west][inner sep=0.75pt]   [align=left] {\begin{minipage}[lt]{94.98pt}\setlength\topsep{0pt}
\begin{center}
{\Large Cases, Deaths}
\end{center}

\end{minipage}};
% Text Node
\draw (799.4,205.22) node [anchor=north west][inner sep=0.75pt]   [align=left] {\begin{minipage}[lt]{183.14pt}\setlength\topsep{0pt}
\begin{center}
{\Large Recoveries, Hospitalizations}
\end{center}

\end{minipage}};
% Text Node
\draw (925.4,275.11) node [anchor=north west][inner sep=0.75pt]   [align=left] {\begin{minipage}[lt]{84.92pt}\setlength\topsep{0pt}
\begin{center}
{\Large Vaccinations}
\end{center}

\end{minipage}};
% Text Node
\draw (758.4,308.6) node [anchor=north west][inner sep=0.75pt]   [align=left] {{\huge \textbf{U.S Census Bureau Data }}};
% Text Node
\draw (812.4,380.7) node [anchor=north west][inner sep=0.75pt]   [align=left] {\begin{minipage}[lt]{170.92pt}\setlength\topsep{0pt}
\begin{center}
{\Large Population Demographics}
\end{center}

\end{minipage}};
% Text Node
\draw (813.4,344.7) node [anchor=north west][inner sep=0.75pt]   [align=left] {\begin{minipage}[lt]{172.25pt}\setlength\topsep{0pt}
\begin{center}
{\Large Vaccine Hesitancy Survey}
\end{center}

\end{minipage}};
% Text Node
\draw (780.99,412.6) node [anchor=north west][inner sep=0.75pt]   [align=left] {{\huge \textbf{Government Responses}}};
% Text Node
\draw (775.4,450.7) node [anchor=north west][inner sep=0.75pt]   [align=left] {\begin{minipage}[lt]{202.75pt}\setlength\topsep{0pt}
\begin{center}
{\Large Nationwide/Statewide Policies }
\end{center}

\end{minipage}};
% Text Node
\draw (511.22,589.7) node [anchor=north west][inner sep=0.75pt]   [align=left] {\begin{minipage}[lt]{32.1pt}\setlength\topsep{0pt}
\begin{center}
{\Large URL}
\end{center}

\end{minipage}};
% Text Node
\draw (931.99,484.6) node [anchor=north west][inner sep=0.75pt]   [align=left] {{\huge \textbf{Local News}}};
% Text Node
\draw (925.4,521.7) node [anchor=north west][inner sep=0.75pt]   [align=left] {\begin{minipage}[lt]{85.45pt}\setlength\topsep{0pt}
\begin{center}
{\Large Title, Content}
\end{center}

\end{minipage}};
% Text Node
\draw (961,556.7) node [anchor=north west][inner sep=0.75pt]   [align=left] {\begin{minipage}[lt]{60.7pt}\setlength\topsep{0pt}
\begin{center}
{\Large Pubisher}
\end{center}

\end{minipage}};
% Text Node
\draw (994,592.7) node [anchor=north west][inner sep=0.75pt]   [align=left] {\begin{minipage}[lt]{32.1pt}\setlength\topsep{0pt}
\begin{center}
{\Large URL}
\end{center}

\end{minipage}};
% Text Node
\draw (605.81,47.52) node [anchor=north west][inner sep=0.75pt]   [align=left] {\begin{minipage}[lt]{101.25pt}\setlength\topsep{0pt}
\begin{center}
{\huge \textbf{CoVaxNet}}
\end{center}

\end{minipage}};

\end{tikzpicture}

}
\caption{An overview of \texttt{CoVaxNet}. This figure shows the online-offline data collected.}
\vspace{-0.5cm}
\label{fig::outline}
\end{figure}

%% file: Related work/relate.tex
% ################### Previous vaccine hesitancy ###############

% In response to the COVID-19 vaccine hesitancy, scientific communities in various disciplines have worked together to combat this crisis. 
Previous studies have investigated vaccine hesitancy on vaccine-preventable diseases such as HPV, H1N1, and Flu~\cite{mcree2014hpv, mesch2015social}. However, vaccine hesitancy can vary on different diseases. The SAGE Working Group on Vaccine Hesitancy (WG) suggests that a complex set of behavioral and social factors should be considered to construct vaccine hesitancy determinants.~\cite{macdonald2015vaccine}. %Thomson et al.~\cite{thomson20165as} introduced \textit{5As}, which is a practical taxonomy for five-dimension non-socio-demographic determinants of vaccine uptake.

% ################## COVID vaccine hesitancy (survey data) ###########

% To better understand COVID-19 vaccine hesitancy, a large number of studies have been conducted with survey data~\cite{loomba2021measuring, geldsetzer2020knowledge, romer2020conspiracy, malik2020determinants, lazarus2021global}. For example, Romer et al.~\cite{romer2020conspiracy} surveyed 1,050 US adults, showing that COVID-19 conspiracy beliefs are related to vaccine hesitancy. Another survey~\cite{lazarus2021global} analyzed age, education level, income level, government, and region-related associations with COVID-19 vaccine acceptance across 19 countries. 

% ################# COVID vaccine hesitancy (social media data) #######

To better understand the COVID-19 vaccine hesitancy, previous works have been conducted on online data.%~\cite{puri2020social, deverna2021covaxxy, lyu2022misinformation}. 
~\cite{deverna2021covaxxy} released the first English Twitter dataset and an online dashboard about COVID-19 vaccines.~\cite{di2022vaccineu} presented a large-scale multilingual Twitter dataset about COVID-19 vaccines.~\cite{muric2021covid} collected 1.8 million anti-vaccine tweets. They investigated the online implicit communities on topic network using the Louvain algorithm.~\cite{lyu2021social} adopt a human-guided machine learning framework on social media data to analyze public opinions on COVID-19 vaccines.~\cite{hussain2021artificial} proposed a sentiment analysis framework to understand the public attitude and concerns on mainstream social media toward COVID-19 vaccines.

% ############### COVID vaccine hesitancy (Offline data) #############
Researchers also analyzed the COVID-19 vaccine hesitancy on offline data. %~\cite{mathieu2021global,huang2022correlation, dirago2022covid}. 
The Oxford COVID-19 Government Response Tracker (OxCGRT) involved over 200 volunteers from the Oxford community to collect publicly available global information on 21 indicators of government response, including data on vaccination policies~\cite{hale2021global}.~\cite{mollalo2021spatial} studied the COVID-19 vaccine hesitancy by analyzing the association between vaccination rates and Social Vulnerability Index (SVI) from CDC on a county level.~\cite{oluyomi2021covid} examined the geospatial trends in COVID-19 incidence of the neighborhood-level U.S. Census data.~\cite{guo2021vaccinations} used human mobility data from Google to analyze the relationship of vaccination rates to mobility and new COVID-19 cases.

%% file: Data collection/data.tex
% short introduction of O2O data
%As shown in Figure~\ref{fig::outline}, in this section we introduce the COVID-19-vaccine online and offline data collected in this repository, including social media data, fact-checking data, COVID-19 statistics, U.S. Census Bureau data, government responses, and local news. Besides, we describe our methodology for data collection, processing and enrichment.
%For the online data we have the following datasets:
% \begin{itemize}[leftmargin=*]
%     \item[$\bullet$] Online Data 
%     \item[$\bullet$] Offline D
% \end{itemize}

\subsection{Data in the Digital World (Online Data)} 
\hypertarget{subsec::onlinedata}{}
\textbf{Tweets:} \label{subsec::onlinedata}
%The scale, volume, and diversity of social media data entail a need for employing advanced crawling and sampling techniques. 
%To collect data that can represent the pro-COVID-19-vaccine and anti-COVID-19-vaccine discourse on Twitter, 
Inspired by~\cite{di2022vaccineu}, we utilize a snowball sampling method in our crawling strategy.
First, we start with collecting publicly available tweets with the two most relevant keywords, \textit{covid} and \textit{vaccine}~\cite{deverna2021covaxxy}, as the initial seeds using Tweepy API\footnote{\url{https://www.tweepy.org/}}. Next, we run n-gram and tf-idf on the collected tweets to identify the most representative and frequent keywords and hashtags for pro-COVID-19-vaccine and anti-COVID-19-vaccine stances. Then, we collect more tweets with newly added keywords and hashtags to expand the list of seeds. We repeat this process until no new seeds can be extracted. %the remaining candidate keywords and hashtags are insufficient and irrelevant for collecting pro-COVID-19-vaccine and anti-COVID-19-vaccine tweets. 
We also carefully analyze the sampled tweets for each keyword and hashtag to check whether they are labeled correctly. For example, \#\textit{Antivaxxers} is initially considered as an anti-vaccine hashtag. However, we find that the sampled tweets are primarily posted by pro-vaxxers to criticize anti-vaxxers. Therefore, we eventually categorize \#\textit{Antivaxxers} as one of the pro-vaccine search keywords. Finally, we remove keywords and hashtags that are not closely related to the COVID-19 vaccine. %such as \textit{Omicron}, \textit{MaskUp}, and \textit{\#NoMasksInClass}.
%  Our Twitter dataset can provide comprehensive, concentrated, and thorough insights into the COVID-19 vaccine-hesitancy with these strategies.

As shown in Table~\ref{tab:keywords}, the shortlisted keywords and hashtags are well-classified to collect COVID-19 pro-vaccine and anti-vaccine tweets. A complete list of them is available in our data repository, which includes 25 pro-vaccine seeds and 28 anti-vaccine seeds. In agreement with Twitter’s Terms of Service, we only release the \textit{tweets IDs} which can be used to retrieve more objects via APIs.  %In the end, we obtained more than 1.49 million English pro-COVID-19-vaccine tweets and 335k anti-COVID-19-vaccine tweets from 555k unique users from January 01, 2020, to January 01, 2022.
\vspace{-0.2cm}
\begin{table*}[!tbh]
    \centering
    \caption{Keywords and hashtags for collecting pro- and anti-COVID-19-vaccine tweets}
    \begin{tabularx}{\textwidth}{|c|>{\centering\arraybackslash}X|}
        \hline
         \textbf{Stance} & \textbf{Keywords \& Hashtags} \\
         \hline
         Pro-vaccine & \textit{get vaccinated, vaccine mandate, vaccination work, fully vaccinated, vaccines save lives, vaccinated for covid, \#GetVaccinatedOrGetCovid, \#Antivacinidiots, \#Antivaxxers, \#GetYourBooster, ...} (25) \\
         
         \hline
         Anti-vaccine & \textit{no vaccine, f**k vaccines, no forced vaccines, no vaccine mandates, \#NoVaccineForMe, \#CovidVaccineIsPoison, \#NoVaccinePassports, \#StopVaccination, \#VaccineSideEffects, ...} (28) \\         
         \hline
    \end{tabularx}
    \label{tab:keywords}
\end{table*}
\vspace{-0.3cm}
\hypertarget{subsec::profiles}{\\}
\textbf{Twitter User Profiles:} \label{subsec::profiles}
We collect metadata from Twitter user profiles, as listed in Figure~\ref{fig::outline}. %include \texttt{usernames}, the number of \texttt{followings}, \texttt{followers}, \texttt{listed memberships}, \texttt{favourites}, \texttt{locations}, \texttt{account creation dates}, and \texttt{profile descriptions (bio)} and \texttt{photos}. 
%Following a previous study~\cite{huang2019large}, 
Due to the lack of geotagged tweets, we leverage the self-reported \texttt{locations} in the users' profiles to tag non-geotagged tweets for data enrichment. Observing that 70.3\% of the collected tweets contain either tweet-level or user-level geo-location information. Furthermore, we use GeoPy~\cite{geopy} to convert the unstructured self-reported \texttt{locations} to structured geographical places with coordinates. %Inspired by recent works~\cite{jung2018assessing, shu2018understanding}, 
We also employ state-of-the-art computer vision and natural language processing tools to infer the demographics of social media users. We use OpenCV~\cite{bradski2008learning} and DeepFace~\cite{taigman2014deepface} to detect human faces and predict age, gender, and race on the \texttt{profile images}. For textual features such as \texttt{usernames} and \texttt{profile descriptions/bio}, by applying Nameparser~\cite{nameparser} and Ethnicolr~\cite{sood2018predicting}, we extract human names to infer race and gender.
% \subsubsection{Geo-locations:}
% As previous studies suggested~\cite{graham2014world, huang2019large}, only around 2 percent of tweets are geotagged. Thus, we supplemented the self-reported locations in the user profiles. Observing that 70.3 percent of the collected tweets contained either tweets-level or users-level geo-location information. Moreover, we used the GeoPy\footnote{\url{https://geopy.readthedocs.io/en/stable/}} library to calculate the coordinates of the locations for each user. 
% \subsubsection{Demographics:}
% In recent years, computational social science is increasingly employing state-of-the-art computer vision and natural language processing tools for different tasks~\cite{jung2018assessing, shu2018understanding, zhang2021influence}. In this study, we used OpenCV~\cite{bradski2008learning} and DeepFace~\cite{taigman2014deepface} to detect human faces and predict age, gender, and race on users' \texttt{profile images}. For text data such as \texttt{usernames} and \texttt{descriptions}, by applying Nameparser\footnote{\url{https://nameparser.readthedocs.io/en/latest/}} and ethnicolr~\cite{sood2018predicting}, we were able to extract the first and last name to predict the race and gender.
\\
\hypertarget{subsec::factcheck}{\\}
\textbf{Fact-checking Reports and Malicious URLs:} \label{subsec::factcheck}
Fact-checking reports can illustrate how COVID-19-vaccine-related mis- and disinformation, rumors, and conspiracy theories spread online. In this dataset, we collect 4,263 COVID-19 vaccine-related fact-checking reports from \texttt{Poynter}\footnote{\url{https://www.poynter.org/ifcn-covid-19-misinformation/}}, where the journalists carefully check the authenticity of news pieces and social media posts. \texttt{Poynter} provides an international fact-checking network (IFCN) alliance collaborating with 100 fact-checking agencies such as \textit{FactCheck.org}, \textit{PolitiFact}, and \textit{Science Feedback} in more than 70 countries and 40 languages. The fact-checkers labeled the articles and social media posts as \textit{false, partially false, misleading,} or \textit{no evidence} according to the reports. We also obtain 813 low credibility sources from \texttt{Iffy+}\footnote{\url{https://iffy.news/iffy-plus/}} dataset, which merges lists of sites that regularly publish mis- and disinformation, as identified by major fact-checking and journalism organizations. 
% We then use this data to label social media posts that contain news pieces from low-credibility sources.

% \subsubsection{Google Trends and Google News:}
% \texttt{Google Trends} and \texttt{Google News} provide insights into the users' information consumption and interests. \texttt{Google Trends} analyzes the search query volumes for specific terms across various regions, languages, and times. \texttt{Google News} aggregates news articles from more than 4,500 sites in English and 50,000 news sources worldwide~\cite{das2007google}. In this work, we collect Google Trends and Google News data with keywords \textit{COVID-19 vaccine} in the United States. We filter out news pieces from low credibility sites and finally collect more than 2,300 news articles.

\subsection{Data in the Physical World (Offline Data)}
\hypertarget{subsec::covidstas}{}
\textbf{COVID-19 Statistics:} \label{subsec::covidstas}
We adopt the COVID-19 data repository operated by the \texttt{Center for Systems Science and Engineering} (CSSE) at Johns Hopkins University~\cite{dong2020interactive}. This data repository obtained statewide, nationwide, and global streaming COVID-19 statistics from reliable sources, including daily confirmed cases, deaths, recoveries, hospitalizations, and vaccinations since January, 2020.%A concise visualization of statewide COVID-19 statistics is partially available in our toolbox \texttt{EXPO2O}\footnote{\url{https://datastudio.google.com/reporting/9ffae165-01d2-4df9-942b-db9ff8889eeb}}.  
\\
\hypertarget{subsec::census}{\\}
\textbf{U.S. Census Bureau Data:} \label{subsec::census}
We comprise the historical state-level and 5-digit zip code-level census records from the \texttt{U.S. Census Bureau}\footnote{\url{https://data.census.gov/cedsci/}} such as the population's age, gender, race, income, health insurance coverage, and employment status. Moreover, the \texttt{Household Pulse Survey}%\footnote{\url{https://www.census.gov/programs-surveys/household-pulse-survey/data.html\#phase3.4}}
(HPS), which focused on estimating how people's lives have been impacted by the COVID-19 pandemic, is available in our data repository. It is worth noting that it consists of survey questions such as \textit{``reasons for children ages 5-17 and adults not receiving or planning to receive a COVID-19 vaccine''}, which is valuable for COVID-19 vaccine hesitancy research. Besides, the COVID-19 Vaccination Tracker from the Census Bureau is also included for a high-level data visualization.
\\
\hypertarget{subsec::responses}{\\}
\textbf{Government Responses:} \label{subsec::responses}
During the COVID-19 vaccine campaign, government policies and responses play an essential role. We obtain the COVID-19-vaccine-related federal and state policies from \texttt{Ballotpedia}\footnote{\url{https://ballotpedia.org/Documenting_America\%27s_Path_to_Recovery}}. It contains spatio-temporal information about COVID-19 vaccine authorization, distribution, and legislation.
\\
\hypertarget{subsec::localnews}{\\}
\textbf{Local News:} \label{subsec::localnews}
We use the GoogleNews~\cite{Googlenews} library to collect real-world events and local news pieces from \texttt{Google News}, which aggregates news articles from more than 4.5k sites in English and 50k news sources worldwide~\cite{das2007google}. In this study, after filtering out low-credibility sources, we acquired around 4.8k news reports with keyword \textit{COVID-19 vaccine} from Jan 2020 to Jan 2022.

%% file: Analysis/analysis.tex
%In this section, we focus on the descriptive analysis of our data repository. We provide preliminary statistics and insights for potential uses of various online and offline data, whereas we leave more sophisticated analysis for future study.
\subsection{Social Media Data}
% \vspace{-0.6cm}
\begin{table*}[tbh!]
    \centering
    \caption{Twitter data statistics}
    \begin{tabularx}{\linewidth}{|>{\centering\arraybackslash}X|>{\centering\arraybackslash}X|>{\centering\arraybackslash}X|}
        \hline
         {\backslashbox{\textbf{ \,\, Features\, }}{\textbf{Stances\,\, }}} & \textbf{Pro-vaccine} & \textbf{Anti-vaccine}\\
         \hline
         Time period & 01.2020 - 01.2022 & 01.2020 - 01.2022\\
         \hline
         \texttt{\#} Tweets & 1,495,991 & 335,229 \\
         \texttt{\#} Retweets, Replies & 317,683/433,465 & 101,059/73,369 \\
         \texttt{\#} Users & 480,327 & 87,405 \\
         
         \texttt{\#} Tweets with URL & 233,673 & 48,635 \\
         \texttt{\#} Tweets with image & 283,365 & 43,127 \\
         \texttt{\#} Tweets with video\texttt{/}GIF & 310,699 & 55,550 \\
         \hline
         Avg \texttt{\#} tweets per user & 3.11 & 3.84 \\
         Avg \texttt{\#} followings per user & 1,354 & 1,473 \\
         Avg \texttt{\#} followers per user & 4,590 & 3,392 \\
         Avg \texttt{\#} favourites per user& 1,374 & 1,506 \\
         Avg \texttt{\#} lists per user& 36.2 & 28.1 \\
         Avg accounts age (years) &  7.91 & 6.55 \\
         %Avg users age (years) & 32.2 & 32.5 \\
        %  Avg users gender (0 \texttt{=} female, 1 \texttt{=} male) & 0.66 & 0.71 \\
         \hline
         Avg \texttt{\#} replies per tweet & 1.08 & 1.42 \\
         Avg \texttt{\#} retweets per tweet & 2.73 & 6.53 \\      
         Avg \texttt{\#} likes per tweet & 13.04 & 21.09 \\      
         Avg \texttt{\#} words per tweet & 25.67 & 23.67 \\      
         \hline
    \end{tabularx}
    \label{tab:compare}
\vspace{-0.3cm}
\end{table*}
% \vspace{-0.6cm}
\begin{figure*}[!tbh]
    \centering
%    \subfigure[Provax Tweets Word Count]{
%    \includegraphics[width=0.47\linewidth]{figure/provax_words.png}
%    }
%    \subfigure[Antivax Tweets Word Count]{
%    \includegraphics[width=0.47\linewidth]{figure/antivax_words.png}
%    }
    \subfigure[Provax Tweets Word Cloud]{
    \includegraphics[width=0.47\linewidth]{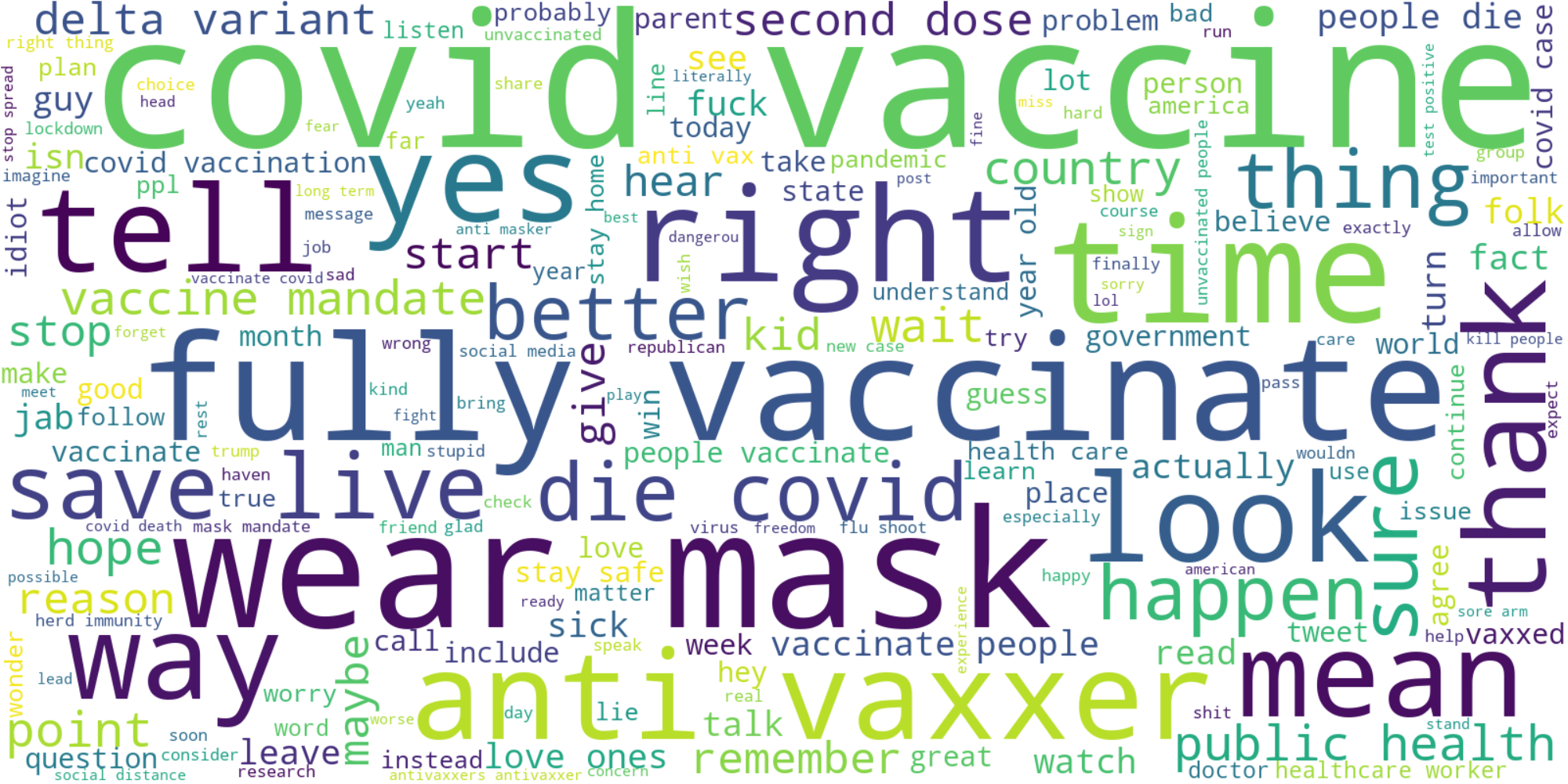}
    }
    \subfigure[Antivax Tweets Word Cloud]{
    \includegraphics[width=0.47\linewidth]{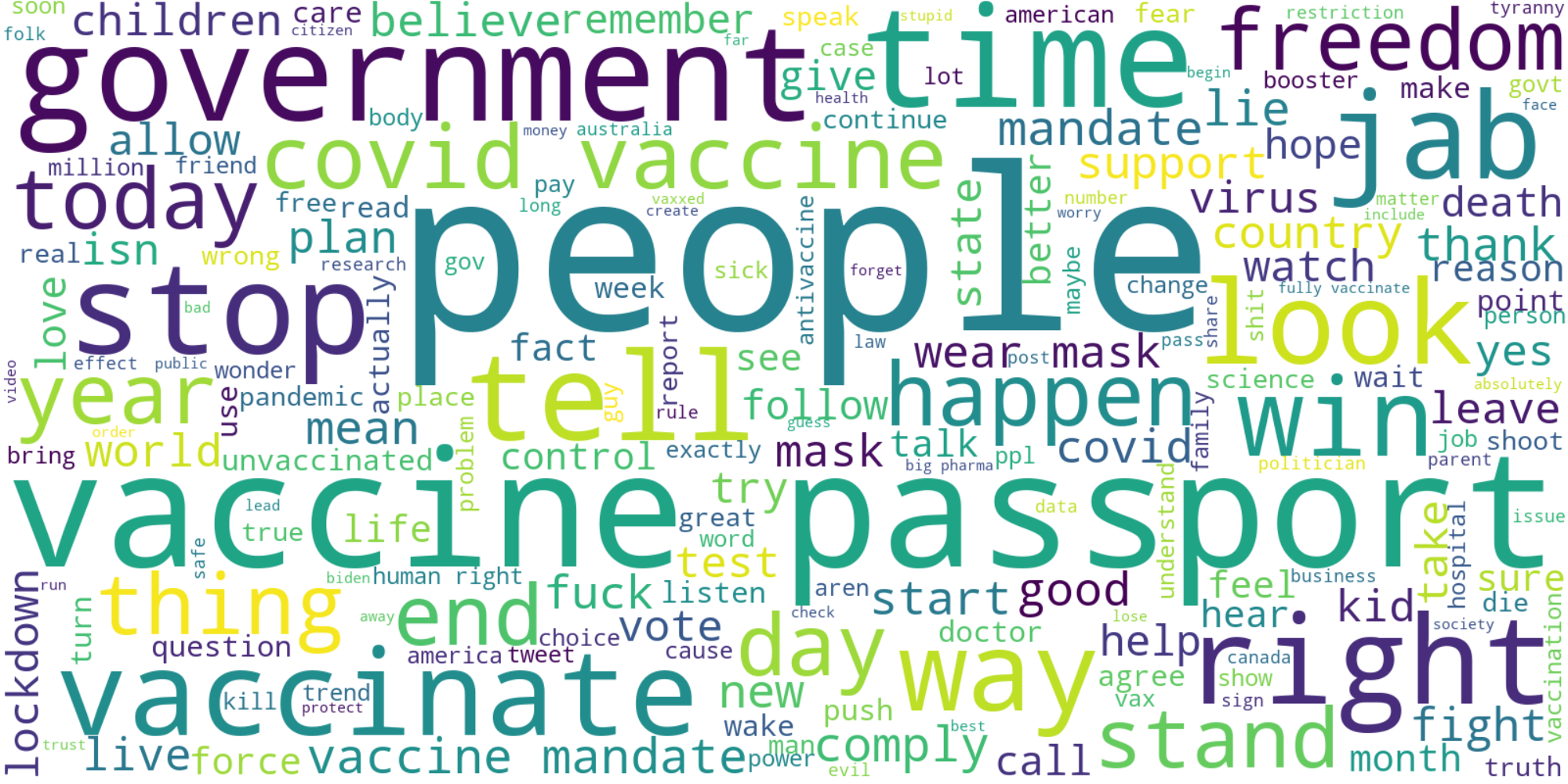}
    }
    \caption{Word clouds of pro- and anti-vaccine tweets. The font size scaled to the frequency.}
    \label{fig:wordcountcloud}
    \vspace{-0.3cm}
\end{figure*}
% \vspace{-0.8cm}
\noindent
The general statistics on our Twitter dataset are presented in Table~\ref{tab:compare}. The Twitter dataset contains 1,831,220 tweets, 366,276 (\texttt{$\sim$}20\texttt{$\%$}) of which have both textual and visual (images or videos) information for multi-modal studies. Figure~\ref{fig:wordcountcloud} reveals the textual characteristics within pro- and anti-COVID-19-vaccine tweets. In the pro-vaccine group, a sufficiently large number of tweets encourage people to get the COVID-19 vaccines and wear masks to save lives. Thus, we can observe that the words frequently appear in this group include \textit{fully, vaccinate, wear, mask, save}. The frequent words from the anti-COVID-19-vaccine tweets, such as \textit{people, stop, freedom, government, passport}, show concern about excessive government control and people's freedom during the pandemic. %On the other hand, both words counts follow a long-tail distribution. The average number of words of pro-COVID-19-vaccine tweets (\texttt{$\sim$}25.67) is slightly greater than anti-COVID-19-vaccine tweets (\texttt{$\sim$}23.67). %Moreover, men are likely to participate more in COVID-19-vaccine online discussions than women. 
\subsection{Fact-checking and Local News Data}
\begin{figure*}[!tbh]
    \centering
    \subfigure[Timeline]{
    \includegraphics[width=0.47\linewidth]{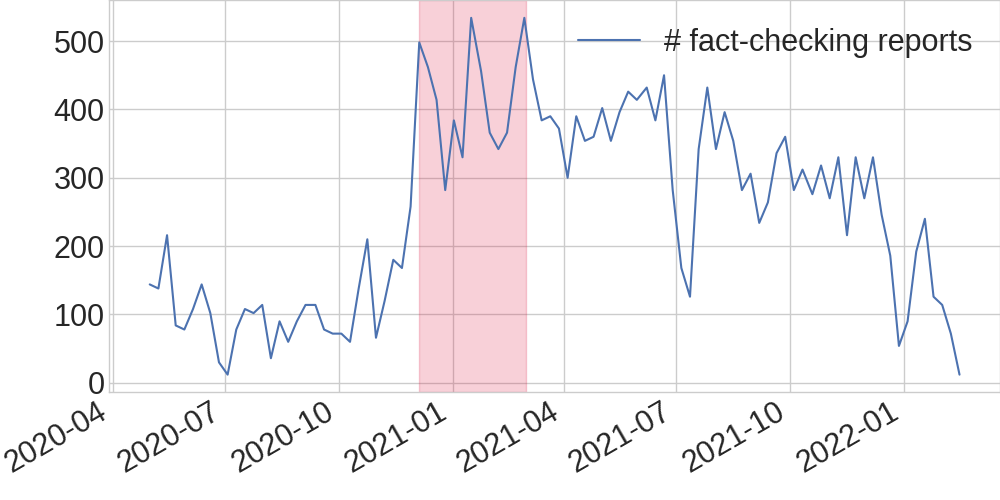}
    }
    \subfigure[Top Fact-checking Websites]{
    \includegraphics[width=0.48\linewidth]{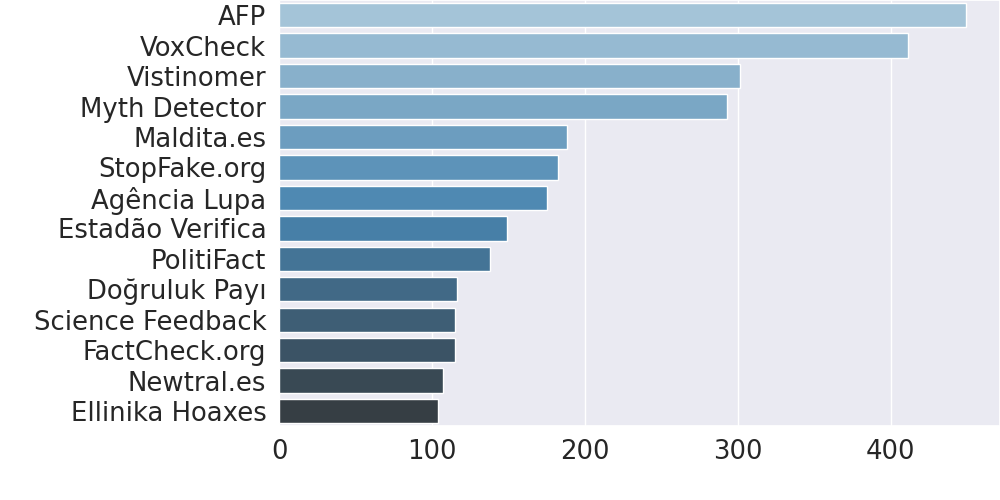}
    }
    \caption{Number of weekly fact-checking reports and top publishers.}
    \label{fig:factcheckstats}
\vspace{-0.6cm}
\end{figure*}

\begin{figure*}[!tbh]
    \centering
    \subfigure[Word Cloud of News Articles]{
    \includegraphics[width=0.47\linewidth]{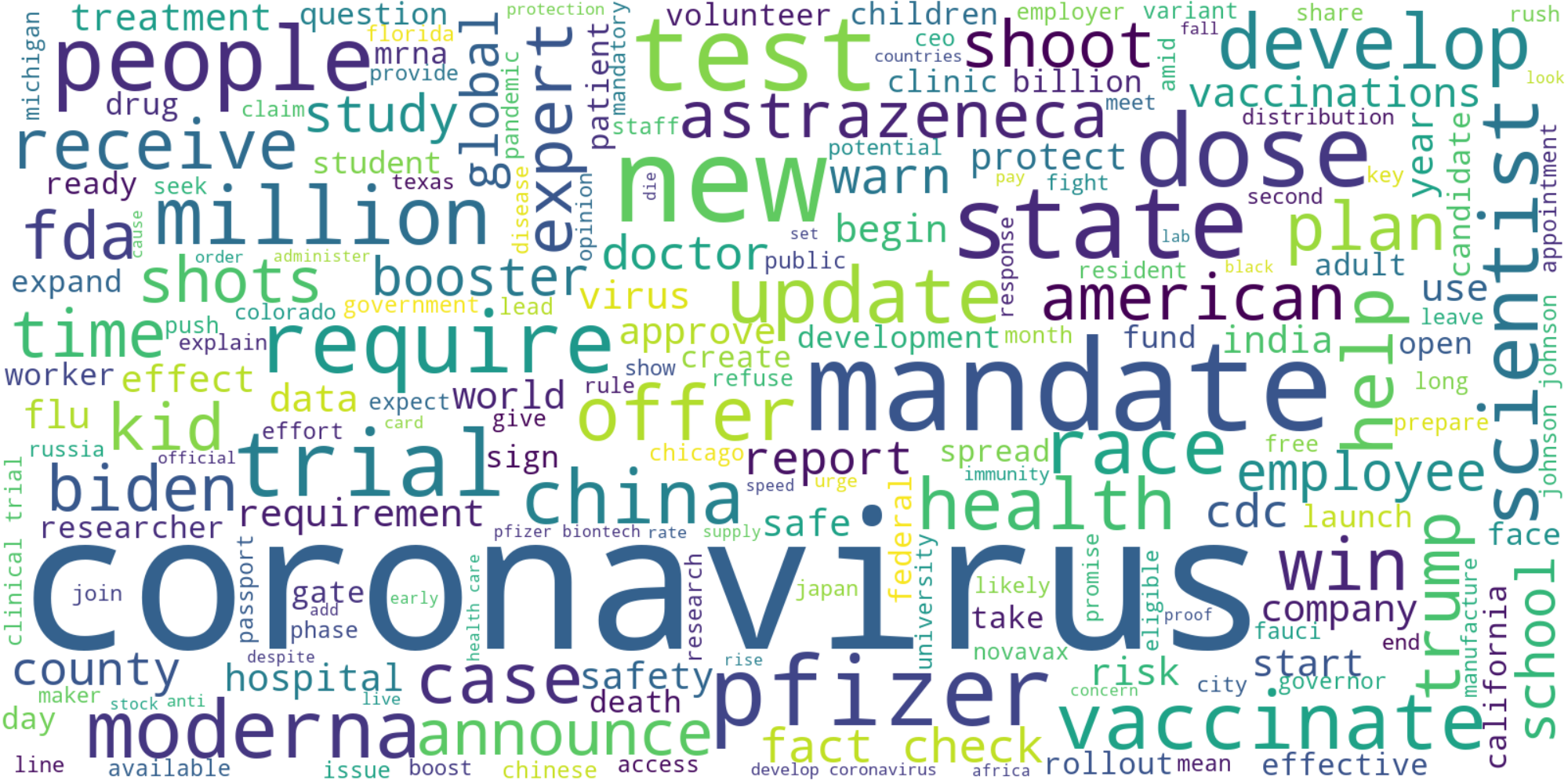}
    }
    \subfigure[Top News Publisher]{
    \includegraphics[width=0.47\linewidth]{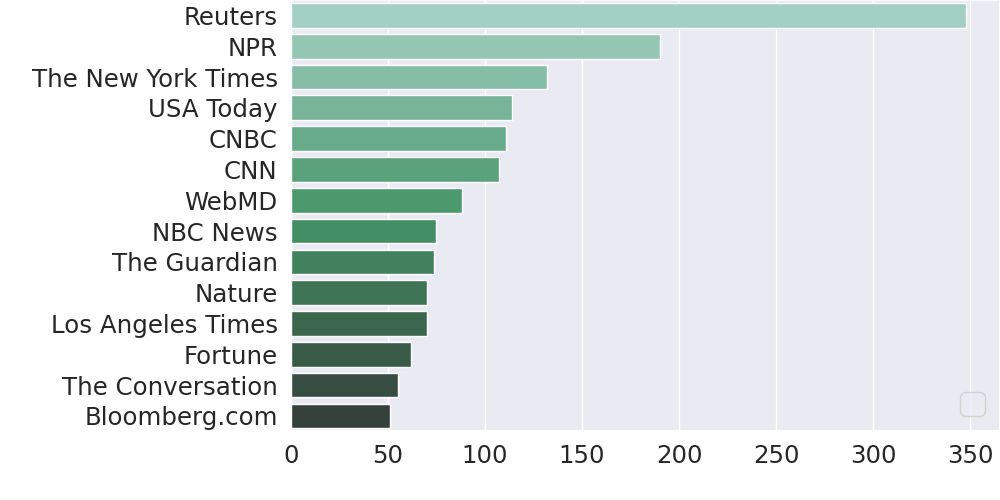}
    }
    \caption{Insights of local news articles.}
    \label{fig:googlenewsstats}
    \vspace{-0.8cm}
\end{figure*}

Figure~\ref{fig:factcheckstats} shows the number of weekly fact-checking reports about COVID-19 vaccines from \texttt{Poynter} and the distribution of active fact-checking websites. We can observe that the number of COVID-19 fact-checking articles published is significantly increased from December 2020 to March 2020, highlighted with a pink rectangle. Note that the emergency use authorization (EUA) approved the first COVID-19 vaccine from Pfizer and BioNTech on December 11, 2020, which potentially led to a massive spread of misinformation about vaccine safety, effectiveness, and adverse reactions. Furthermore, this dataset contains fact-checking websites from various countries, which provides an potential for multilingual misinformation studies on COVID-19 vaccines. 
Figure~\ref{fig:googlenewsstats} presents the word cloud and top news publishers of the local news dataset. We can observe that news authors' frequently used topics and vocabularies are more neutral and objective than those of Twitter users. A possible explanation is that the news articles collected are from multiple sources, including cross-domain for-profit and non-profit news outlets, multidisciplinary science journals, and international health magazines.
\vspace{-0.7cm}
\vspace{-0.8cm}
% \vspace{-0.2cm}
\subsection{Offline Statistics}
%Figure~\ref{fig:vaccinetrend} 
In the repository, CSV files are well-documented with sufficient offline statistics. The JHU COVID-19 dashboard\cite{dong2020interactive} provides easy-access, interactive, and comprehensive data visualization of spatio-temporal COVID-19 statistics. Note that COVID-19 statistics and Census data have shown to be correlated with vaccine hesitancy in a quantitative study\cite{lyu2022misinformation}. Similar findings from surveys can be good validations. For example, \texttt{Kaiser Family Foundation} (KFF) compares public attitudes towards COVID-19 vaccines among different demographic and socioeconomic statuses through a monthly survey. We can observe that: (i) political lean is correlated with vaccine attitudes $-$ Republicans tend to refuse the COVID-19 vaccine more than Democrats; (ii) The vaccine attitudes are similar among different races across time; (iii) People in rural areas are more likely to be anti-vaxxers than those in suburban and urban areas; (iv) People with a college degree are more willing to vaccinate than those with less than a college degree; and (v) The percentage of pro-vaxxers is increasing while anti-vaxxers' remains stable and low over time among all demographic and socioeconomic groups.
% \begin{figure*}[!tbh]
%     \centering
%     \subfigure[Vaccine Attitudes by Political Leans]{
%     \includegraphics[height = 0.38\linewidth, width=0.47\linewidth]{figure/party-id-april.png}
%     }
%     \subfigure[Vaccine Attitudes by Races]{
%     \includegraphics[height = 0.38\linewidth, width=0.47\linewidth]{figure/race-april.png}
%     }
%     \subfigure[Vaccine Attitudes by Communities]{
%     \includegraphics[height = 0.38\linewidth, width=0.47\linewidth]{figure/usr-april.png}
%     }
%     \subfigure[Vaccine Attitudes by Education Levels]{
%     \includegraphics[height = 0.38\linewidth, width=0.47\linewidth]{figure/education-april.png}
%     }
%     \caption{Vaccine attitude among different demographic and socioeconomic status.}
%     \label{fig:vaccinetrend}
% \end{figure*}

%% file: Connection/connection.tex
The COVID-19 vaccine campaign gave rise to a surplus of online and offline data expressing people's views and different policies imposed by the government to promote the vaccine. These two kinds of data share a strong relationship. For example, offline COVID-19 statistics can alter the vaccine stance of the online crowd~\cite{lyu2021social}. Thus, online and offline data can complement each other to optimize the information capture. However, connecting online and offline data is challenging because they reside in different sources, modalities, and dimensions. To this end, we first discuss how to connect COVID-19 online and offline data in \texttt{CoVaxNet}. Furthermore, we conduct COVID-19 vaccine stance detection to (i) verify the quality and compatibility of our datasets for various downstream tasks, including stance detection; (ii) demonstrate the necessity for employing online-offline data; and (iii) encourage researchers to develop state-of-the-art models and incorporate various online and offline data provided by \texttt{CoVaxNet} to facilitate the COVID-19 vaccine hesitancy study.
\vspace{-0.2cm}
\subsection{Data Connection}
An online-offline dataset contains (i) an online dataset and (ii) an offline dataset. Thus, we denote the online-offline dataset $\mathcal{D}_{online-offline}$ as
\vspace{-0.2cm}
\begin{equation} \label{eq:O2O_eq}
\mathcal{D}_{online-offline} = (\mathcal{D}_{online},\, \mathcal{D}_{offline}),
\vspace{-0.2cm}
\end{equation}
where~$\mathcal{D}_{online} = \{{d}_{online}^{1}, \, {d}_{online}^{2}, \, ..., \, {d}_{online}^{n}, \, ..., \, {d}_{online}^{N}\}$ denotes the set of $N$ online data and $\mathcal{D}_{offline} = \{{d}_{offline}^{1}, \, {d}_{offline}^{2}, \, ..., \, {d}_{offline}^{m}, \, ..., \, {d}_{offline}^{M}\}$ is the set of $M$ offline data. Then an online-offline data connection $\mathcal{C}$ which establishes a link between the online and offline dataset, presented by implicit or explicit features, can be extracted. In this work, we connect online social media dataset $\mathcal{D}_{online}$ with offline COVID-19 statistics dataset $\mathcal{D}_{offline}$ through a geolocation-based online-offline connection $\mathcal{C}_{geo}$. For example, given two online social media posts $\mathcal{P}_{1}$ and $\mathcal{P}_{2}$ from $\mathcal{D}_{online}$ and two offline COVID-19 statistics $\mathcal{S}_{1}$ and $\mathcal{S}_{2}$ from $\mathcal{D}_{offline}$, and a connection $\mathcal{C}_{geo}$ such that 
\vspace{-0.1cm}
\begin{gather*} \label{O2O_eq_geo1}
\mathcal{C}_{geo}(\mathcal{P}_1) = \mathcal{C}_{geo}(\mathcal{S}_1) = \{\textit{Tempe, AZ, USA}\}\\
\mathcal{C}_{geo}(\mathcal{P}_2) = \mathcal{C}_{geo}(\mathcal{S}_2) = \{\textit{Pittsburgh, PA, USA}\}.
\vspace{-0.2cm}
\end{gather*}
Therefore, the online-offline dataset $\mathcal{D}_{online-offline}$, can be formulated as
\vspace{-0.2cm}
\begin{equation} \label{O2O_eq_geo2}
\begin{split}
\mathcal{D}_{online-offline} & = (\mathcal{D}_{online},\, \mathcal{D}_{offline}) \\
 & = \{<\mathcal{P}_{1}, \mathcal{S}_{1}>,\, <\mathcal{P}_{2}, \mathcal{S}_{2}>, \, ...\},
\end{split}
\vspace{-0.2cm}
\end{equation}
where the tuple $<\mathcal{P}_{1}, \mathcal{S}_{1}>$ indicates the post $\mathcal{P}_{1}$ and the statistics $\mathcal{S}_{1}$ are linked to each other through the geolocation indicator $\mathcal{C}_{geo}$.

\vspace{-0.2cm}
\subsection{COVID-19 Vaccine Stance Detection}
We adopt $\mathcal{D}_{online}$, $\mathcal{D}_{offline}$, and $\mathcal{D}_{online-offline}$ in the last subsection to perform vaccine stance detection. For data preprocessing, we encode the tabular COVID-19 statistics into text representations so as to make them consistent with the modality of tweets. 
First, we compute the weekly COVID-19 severity of each city
\vspace{-0.2cm}
\begin{equation} \label{eq:O2O_eq2}
Severity = \sum_{i=1}^{7} (\mathcal{G}_{i}^{c} + \mathcal{G}_{i}^{d} + \mathcal{G}_{i}^{h} - \mathcal{G}_{i}^{r}-\mathcal{G}_{i}^{v}),
\vspace{-0.2cm}
\end{equation}
where $i$ is the number of days before a tweet is posted. $\mathcal{G}_{i}^{c}$, $\mathcal{G}_{i}^{d}$, $\mathcal{G}_{i}^{h}$, $\mathcal{G}_{i}^{r}$, and $\mathcal{G}_{i}^{v}$ denote the daily growth rate of \textit{cases, deaths, hospitalizations, recoveries,} and \textit{vaccinations}. Second, for a given list of cities, we categorize the \textit{$<$severity level$>$} of each city as \textit{low}, \textit{medium}, or \textit{high} by sorting the Severity values in an ascending order and then splitting the list based on the proportion of
$0.33:0.33:0.33$. Third, we encode the \textit{$<$severity level$>$} at the end of each tweet. For example, a encoded tweet will be ``{{mRNA is made from Human DNA. It's designed to attach to your DNA.} \textit{The COVID-19 severity of my city is <severity level>}}.'' In addition, we encode the user demographics, including age, gender, and race, from the user profile into the tweets. For instance, "I am a $<$age$>$ years old $<$race$>$ $<$gender$>$." Now we have three types of data: (i) \textit{online data} (tweets); (ii) \textit{offline data} (encoded COVID-19 statistics); and (iii) \textit{online-offline data} (tweets $+$ encoded COVID-19 statistics and user demographics).

We randomly selected 70\% data as the training set and 30\% as the testing set, as well as a balanced distribution ($1:1$) between pro- and anti-vaccine tweets. We deploy four well-established baselines including Support Vector Machines (SVM), Naive Bayes (NB), Logistic Regression (LR), and BERT in our experiments. The experimental results are listed in Table~\ref{tab:ex_result}. We observe that four baselines have reasonably good performance results in terms of F1-score and accuracy. This validates the quality of labels and collected features in \texttt{CoVaxNet}. Moreover, we can find that models trained on online-offline data outperform models trained on either online or offline data. This demonstrates that online-offline data can provide more critical information than online or offline data, which can be learned by the baseline models. However, the experimental results of these baselines are still limited and shows great potential for improvements.
\vspace{-0.5cm}
\begin{table*}[!tbh]
    \centering
    \caption{COVID-19 vaccine stance detection performance on \texttt{CoVaxNet}}
    \vspace{-0.2cm}
    \begin{tabular}{|p{0.2\textwidth}|cc|cc|cc|cc|}
         \hline
         \multicolumn{1}{|c|}{Method} & \multicolumn{2}{c|}{SVM} & \multicolumn{2}{c|}{NB} & \multicolumn{2}{c|}{LR} & \multicolumn{2}{c|}{BERT} \\
         \hline
         \multicolumn{1}{|c|}{Metric} & \multicolumn{1}{c|}{Accuracy} & \multicolumn{1}{c|}{F1} & \multicolumn{1}{c|}{Accuracy} & \multicolumn{1}{c|}{F1} & \multicolumn{1}{c|}{Accuracy} & \multicolumn{1}{c|}{F1} & \multicolumn{1}{c|}{Accuracy} & \multicolumn{1}{c|}{F1} \\
         \hline
         \multicolumn{1}{|c|}{Online} &  \multicolumn{1}{c|}{0.736} & \multicolumn{1}{c|}{0.763}	& \multicolumn{1}{c|}{0.758} & \multicolumn{1}{c|}{0.751} & \multicolumn{1}{c|}{0.776} & \multicolumn{1}{c|}{0.777} & \multicolumn{1}{c|}{0.784} & \multicolumn{1}{c|}{0.788} \\ 
         \multicolumn{1}{|c|}{Offline} & \multicolumn{1}{c|}{0.630} & \multicolumn{1}{c|}{0.602} & \multicolumn{1}{c|}{0.658} & \multicolumn{1}{c|}{0.667} & \multicolumn{1}{c|}{0.674} & \multicolumn{1}{c|}{0.676} & \multicolumn{1}{c|}{0.675} & \multicolumn{1}{c|}{0.680} \\
        %  CNN & 0.88 & 0.88 & 0.79	& 0.78 &  \textbf{0.80} & \textbf{0.80} & 0.75 & 0.73 & 0.82	& 0.81 & \textbf{0.80} & \textbf{0.80}\\
         \multicolumn{1}{|c|}{Online-Offline} & \multicolumn{1}{c|}{\textbf{0.759}} & \multicolumn{1}{c|}{\textbf{0.769}} & \multicolumn{1}{c|}{\textbf{0.782}} & \multicolumn{1}{c|}{\textbf{0.777}} & \multicolumn{1}{c|}{\textbf{0.789}} & \multicolumn{1}{c|}{\textbf{0.788}} & \multicolumn{1}{c|}{\textbf{0.796}} & \multicolumn{1}{c|}{\textbf{0.793}} \\
        %  \midrule 
        %  XGB\_{Social} & & & \\
        %  \midrule
                %  \midrule
        %  XGB\_{Content+Social} & & & \\
        %  TCNN\_URG & 0.86 & 0.81 & 0.91 \\
        %  dEFEND & 0.93 & 0.91 & 0.53 \\
         \hline
    \end{tabular}
    \label{tab:ex_result}
\end{table*}
\vspace{-0.9cm}

%% file: Conclusion/conclusion.tex
To combat the COVID-19 pandemic, we present a multi-source, multi-modal, and multi-feature online-offline repository for COVID-19 vaccine hesitancy research. We introduce our strategy for constructing a high-quality online-offline data repository. We share sufficient insights, which detail textual, visual, spatio-temporal, and network features in the data. Furthermore, we propose a geolocation-based online-offline data generation approach. We believe \texttt{CoVaxNet} would facilitate the research of analyzing the impact of various online and offline data on COVID-19 vaccine hesitancy and help policymakers prioritize resource allocation in the ongoing and future pandemics. There are several promising future directions. First, we will improve our labeling strategy to reduce noise and provide ground truth of \textit{``vaccine hesitancy''} in our Twitter dataset. Second, we intend to extend the \texttt{CoVaxNet} to include more relevant data from other reliable sources. For example, we will collect data from countries with similar and dissimilar COVID-19 vaccine roll-out plans. This will help researchers study the structural inequalities and the correlation among all information regarding vaccine uptake during the pandemic.